\documentclass[12pt,a4paper]{report}
\usepackage{graphicx}
\usepackage{here}
\usepackage{float}
\usepackage[nospace,superscript]{cite}
\usepackage{indentfirst}
\usepackage{geometry}
\usepackage{amsmath}
\usepackage{amsfonts}
\usepackage{color}
\usepackage[french,english]{babel}
\usepackage[utf8]{inputenc}
\usepackage[colorlinks=true]{hyperref}
\hypersetup{urlcolor=black,linkcolor=black,citecolor=black,colorlinks=true}
\usepackage{chngcntr}
\counterwithout{figure}{chapter}
\counterwithout{table}{chapter}
\counterwithout{equation}{chapter}

\begin{document}


\begin{figure}[!t]
\rule[0.5ex]{\textwidth}{0.1mm}
\center
\includegraphics[height=17mm]{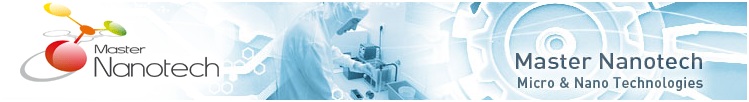}
\label{logo_nano}
\end{figure}


\hfill

\begin{center}
\Large{Master thesis report\\(arXiv research extract only)}\vspace{1.4cm}
\\\Large{\bfseries{Quantum Technology: Single-Photon Source}}
\vspace{1.4cm}

\large{Author:\vspace{0.25cm}\\ Vincent Camus\\ Master Nanotech 2011-2012 - Grenoble INP Phelma\\ $vincent.camus@phelma.grenoble$-$inp.fr$
\vspace{1.4cm}

Supervisors:\vspace{0.25cm}\\Prof. Kae Nemoto\\NII - 2-1-2 Hitotsubashi, Chiyoda-ku, Tokyo 101-8430 Japan\\ $nemoto@nii.ac.jp$\vspace{0.12cm}\\Simon Devitt\\NII - 2-1-2 Hitotsubashi, Chiyoda-ku, Tokyo 101-8430 Japan\\ $devitt@nii.ac.jp$
\vspace{1.4cm}

Defended in September 2012\\
\vspace{0.05cm}Declassified in February 2017}
\end{center}


\begin{figure}[!b]
\center
\includegraphics[height=17mm]{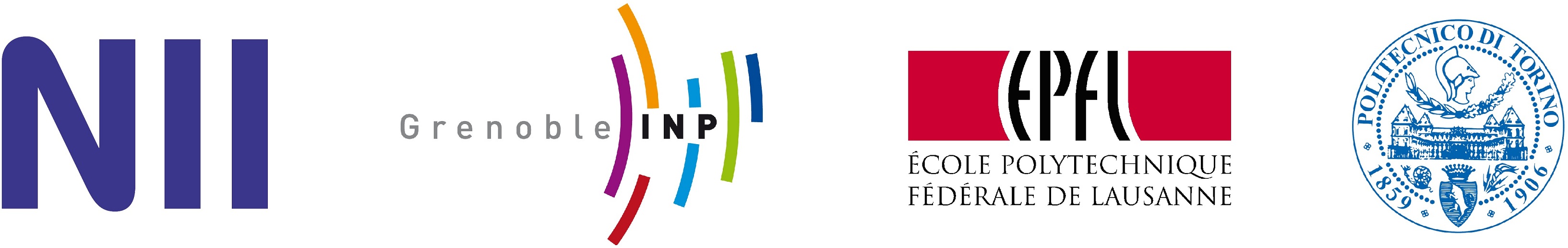}
\rule[0.5ex]{\textwidth}{0.1mm} 
\label{logos}
\end{figure}
\newpage




\section*{Abstract}

This report is a synthesis of my master thesis internship at the National Institute of Informatics (NII) in Tokyo, Japan, that lasted during the summer of year 2012. I worked in the Quantum Information Science Theory (QIST) group under supervision of Prof. Kae Nemoto and Dr. Simon Devitt. This group works on theoretical and experimental implementations of quantum information science.\\ 

The aim of my project was to study and improve quantum optical systems. I first studied different fields and systems of quantum information science. Then I focused my research on single-photon sources, entangled photon sources and interferometric photonic switches. Finally, I found some strategies to design an efficient and optimized single-photon source that could be built with today's technologies. This report describes in details the created and optimized design of a single-photon source based on time and space multiplexing of Spontaneous Parametric Downconversion (SPDC) sources.

\section*{Résumé (French)}

Ce rapport est une synthèse de mon stage de fin de master à l'Institut National d'Informa- tique de Tokyo, qui s'est déroulé pendant l'été 2012. J'ai travaillé dans le groupe de recherche d'information quantique (QIST) sous la supervision du Prof. Kae Nemoto et du Dr. Simon Devitt. Ce groupe travaille à la réalisation théorique et pratique de systèmes et modules quantiques.\\

Le but de mon projet était d'étudier et de chercher à améliorer les systèmes optiques quantiques. J'ai tout d'abord étudié plusieurs domaines et systèmes relatifs à la théorie de l'information quantique. Par la suite, je me suis focalisé sur les sources monophotoniques, les sources photoniques intriquées et l'interférométrie quantique. Enfin, j'ai trouvé une nouvelle stratégie pour améliorer et optimiser des sources monophotoniques implémentables avec les technologies d'aujourd'hui et de demain. Ce rapport décrit en détail le fonctionnement et les composants d'une source basée sur le multiplexing spacial et temporel de sources de type SPDC (Spontaneous Parametric Downconversion).

\section*{Riassunto (Italian)}

Questo documento è una sintesi del mio stage di master, effettuato presso l'Istituto Nazionale di Informatica nel Tokyo nell'estate 2012. Ho lavorato nel gruppo di ricerca di informazioni quantistiche (QIST) sotto la supervisione del Prof. Kae Nemoto e del Dr. Simon Devitt. Questo gruppo lavora nella realizzazione di moduli teorici e pratici per sistemi quantistici.\\

L'obiettivo del mio progetto era di studiare e migliorare i sistemi ottici quantistici. Nella prima parte del mio lavoro, ho studiato vari domini e sistemi, inerenti alla scienza dell'informa- zione quantistica. Successivamente, ho focalizzato la mia attenzione sulle sorgenti di singoli fotoni, fotoni intricati, e, allo stesso tempo, interferometria quantistica. Infine, ho trovato delle nuove strategie al fine di migliorare e ottimizzare le sorgenti di singoli fotoni, le quali potrebbe essere implementate con le tecnologie attuali. Questo documento descrive in dettaglio il funzionamento e i componenti delle sorgenti basate sulla moltiplicazione spazio-temporale SPDC (Spontaneous Parametric Downcoversion).

%
%
%


\tableofcontents

\chapter{Spontaneous Parametric Downconversion single-photon source}

\section{Single-photon source}

An ideal single-photon source\cite{sps} is a source that can emit at any arbitrary and defined time (\emph{on-demand}) and in which the probability of  single photon emission is 100\% and the probability of multiple-photon emission is 0\%. Emitted photons are indistinguishable, repetition rate is arbitrarily fast. Deviations from these ideal characteristics.\\

We imagine building single-photon sources from various systems: {quantum dots\cite{spsqd1,spsqd2,spsqd3}}, {single atom\cite{spssa}}, ion\cite{spssi} or {molecule\cite{spssm}}. Another possibility is to use a probabilistic system based on photon Spontaneous Parametric Downconversion\cite{spsspdc1,spsspdc2} (SPDC).

\section{Single-photon detector}

Single-photon detectors based on single-photon avalanche photodiodes, photomultiplier tubes, superconducting nanowires are typically used as \emph{non photon-number resolving} detectors. They can only distinguish zero photon and more than zero photons, and they are the most commonly used single-photon detectors.\\

Others are \emph{photon-number resolving} detectors. While detecting a single photon is a very difficult task, discriminating the number of incident photons is even more difficult. One direct approach is simply to break the detector active area into distinct pixels and split the idler signal onto these pixel areas. Another approach is based on superconducting tunnel junction, but their complexity are high, their efficiencies are not very high, their resolution and speed are limited, and their working temperature very low ($<$~0.4~K).

\section{SPDC theory}

The response of an optical material to incoming fields is the combination of responses from a large number of atoms making up that material. The Maxwell equations for a simple linear material response  wit electric given by the following equation:
\begin{equation}
\textbf{P} = \epsilon_0 \chi \textbf{E}
\end{equation}
where $\textbf{E}$ is the original electric field, $\textbf{P}$ is the the material polarization and $\chi$ the susceptibility tensor of the material (for isotropic media like a gas or a glass, it is a scalar).\\

In a nonlinear material, the response of the medium to an exciting field is non-linear, as:
\begin{equation}
\textbf{P} = \epsilon_0 ( \chi \textbf{E} + \chi^{(2)} \textbf{E}^2 +\chi^{(3)} \textbf{E}^3 ... )
\end{equation}
where $\chi^{(n)}$ are the susceptibility tensors of higher order in the material response. The nonlinear response is usually small. The second order term $\chi^{(2)}$ is due to a lack an inversion symmetry in the material crystal. Thus amorphous materials like glass or polymers present the property $\chi^{(2)} = 0$.\\

For a monochromatic laser, the electric field is of the form:
\begin{equation}
\textbf{E}(x,t) = \textbf{E}_0 e^{\emph{\textbf{i}} ( k x - \omega t)}
\end{equation}
The second order nonlinear susceptibility results in a polarization component which oscillate at twice the original frequency:
\begin{equation}
\textbf{P}^{(2)}(t) \propto \chi^{(2)} \textbf{E}^2 \propto \chi^{(2)} \textbf{E}_0^2 e^{\emph{-2\textbf{i}} \omega t}
\end{equation}
This polarization component at a new frequency can be considered as a source in the Maxwell equations, and will propagate through the material according to the dispersion relation.\\

\section{SPDC source}

\begin{figure}[H]
\center
\includegraphics[scale=0.33]{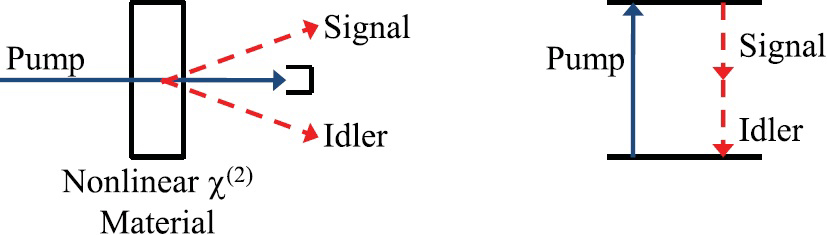}
\caption{Spontaneous Parametric Downconversion of one photon into two photons.}
\label{spdc_theory}
\end{figure}

In a SPDC source, a pump laser illuminates a material with a $\chi^{(2)}$ optical nonlinearity, creating two photons, called Signal and Idler, under the constraints of momentum and energy conservation (Fig.~\ref{spdc_theory}). The detection of one trigger photon indicates, or heralds, the presence of its twin, that can be send into a photon-source output.\\

It is a probabilistic process. Because of this nature, in addition to the probability of generating one pair of photons, there is also a finite probability of generating more than one pair via higher order emissions.\\

Using a pulsed laser to pump the nonlinear crystals, the probability of generating $n$ pairs of photons per pulse is led by a Poisson distribution: 
\begin{equation}
P_n = \frac {N^n e^{-N}} {n!}, n \in \mathbb{N}
\end{equation}
where $N$ is the mean number of pairs, it depends on the pump power and parameters of the crystals.\\

Such behavior decreases the quality of the single-photon source. Since \emph{photon-number resolving} detectors are too complicated to integrate and too expensive, \emph{non photon-number resolving} detectors are used to herald the Idler photon. SPDC heralded sources must be held to average pair production levels much less than one to avoid producing multiple pairs which would result in the heralded channel containing more than one photon.

\section{Optical switch}

\begin{figure}[H]
\center
\includegraphics[scale=0.5]{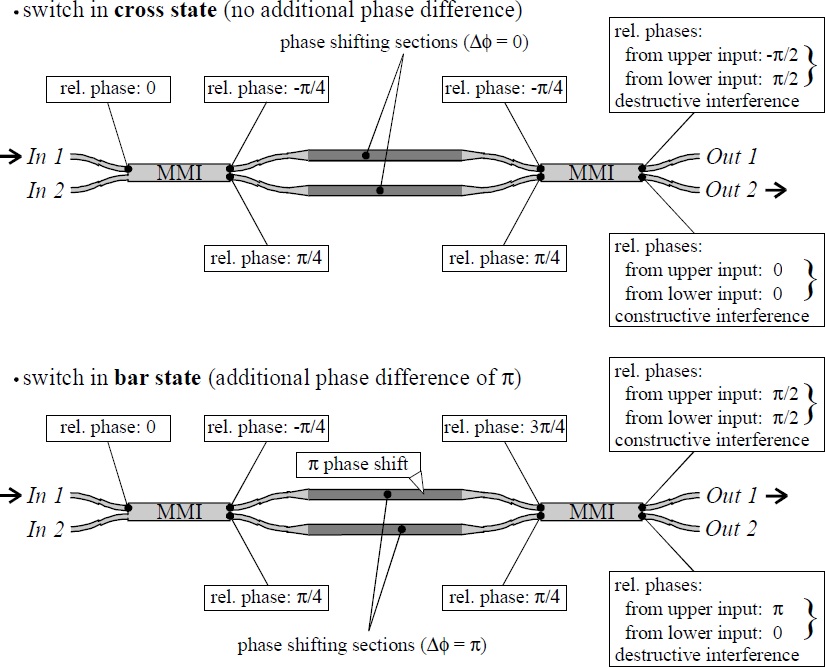}
\caption{Schematic layout of the waveguiding structure of the MZI based {switch\cite{mzi_thesis}}.}
\label{mzi_switch}
\end{figure}

The best optical switch for single photons consists in a Mach-Zehnder Interferometer\cite{mzi_thesis} (MZI). A 3~dB splitter and a~3 dB combiner connected by two interferometer arms. By changing the effective refractive index of one of the arms, the phase difference at the beginning of the combiner can be changed, such that the light switches from one output port to the other. The structure is fabrication tolerant and polarisation insensitive. A schematic layout of the MZI-based switch is depicted on~Fig.~\ref{mzi_switch}.

\section{SPDC based source}

Most of SPDC based sources work with a temporal or a spatial multiplexing, thus have the same frequency as pumping or a (low) divisor. In both cases, the multiplexing cannot control more than one photon and the Poissonian statistics of SPDC leads to a waste of photons or a stronger multi-photon {rate\cite{aa1,aa2,aa3}}$^{ (links)}$. Some SPDC based sources also try to use photon storage into different kinds of loops or cavity systems without meeting great {success\cite{aa4,aa5}}$^{ (links)}$.


\chapter{Design of SPDC based single-photon source with output rate up-conversion multiplexed architecture}
\section{Generalities}

\subsection{Abstract}

We describe a single-photon source based on an array of spontaneous parametric downconversion (SPDC) modules multiplexed by an integrated structure composed of electro-optic switches and delay wave-guides.\\

The suggested optical device allows to increase the output photon source rate compared to the SPDC module laser pump rate and enhances both stability, controllability, and efficiency.\\

This device transforms the spatial multiplexing into a temporal multiplexing with temporary photon storage. The photons are stored into different delay lines and temporally sequentialized. This train of photon can then be used in high frequency emission and compensation of photon lack among the next pump cycles.\\

The implementation consists of a \textbf{crossed multiplexed architecture using electro-optic switches}. The architecture first drives the downconverters array outputs into several layers of \textbf{delay lines} organized in \textbf{shared and by-passable binary delay register}. The architecture then includes a \textbf{routing tree} driving all the photons into a single output.

\subsection{Aim}

I aimed at creating a high efficiency device, eliminating the waste of photons and the stability problems due to probabilistic behavior of spontaneous downconversion. I also wanted a different way of thinking. The tools are known, the direction is different, but the result is there, and better. It takes a logical and optimistic look toward advanced research on quantum computer.\\

The main advantages of the frequency up-conversion toward the pumping rate is the reduction of length of the integrated optics and the possibility to overtake heralding detector problems. While progress of integrated MZI interferometers, integrated wave-guides or photonic crystals is prompt, single-photon detector technology, number-resolving or non-number-resolving, is still far from the same level of development.\\

\textbf{This suggested device offers the possibility to build an advanced integrated optical device, moreover functional element of a future quantum computer, whose design gives the most of its flexibility to its detectors and the most of its potential to its optical control.}


\section{General scheme}

\begin{figure}[H]
\center
\includegraphics[scale=0.7]{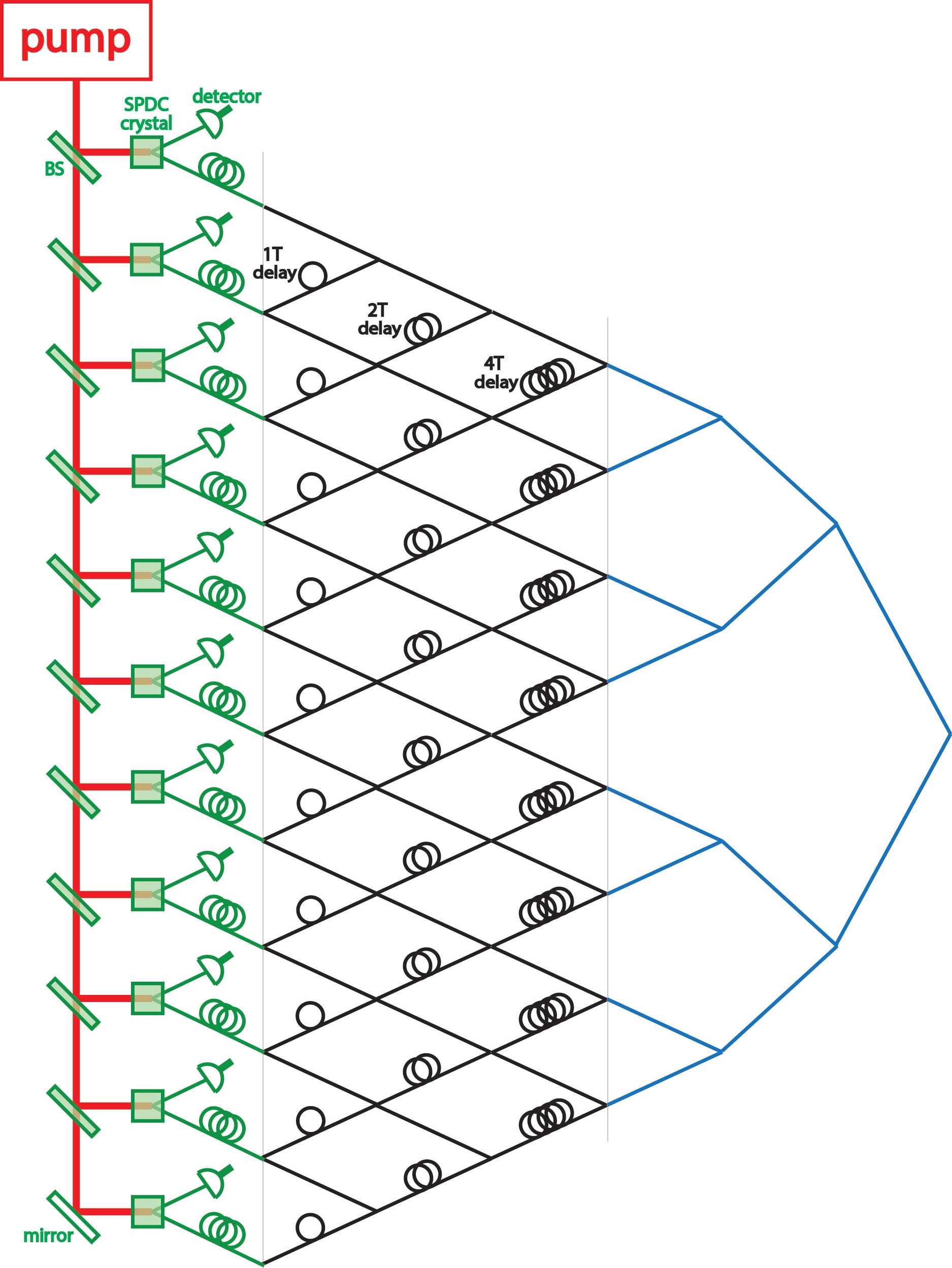}
\caption{General scheme of a single-photon source with 11 SPDC crystals and $0T$ to $7T$ delay register. (red) Pumping laser. (green) Downconverters (pair generator, heralding detector and heralding decision delay). (black) Crossed by-passable delay register. (blue) Routing tree.}
\label{design_exemple11}
\end{figure}


\section{Detailed description of elements}

\subsection{Downconverters}

The pump laser illuminates a array of $\chi^{(2)}$ non-linear crystals, creating both idler and signal photons. The idler photons are sent to heralding detectors connected to a processor unit (FPGA). The signal photons are sent into the integrated architecture through delay fibers to let time for the processor unit to take path decisions.\\

The delay fibers have slightly different lengths to correct the entering time differences among emitted photons due to the growing distance between the pump and each crystal of the array.\\

The number of SPDC modules can be chosen and optimized for the symmetry of the linear then logarithmic routing into the single output. In the hypothesis of a photon driving demonstration, single-photon or multi-photon are driven in the same way through standard MZI switches. Thus, a small number of downconverters (below twenty) can be chosen with an exaggerated photon rate. For a demonstration of efficient single-photon source, number of downconverters must be higher (from a few tens to more than hundred).


\subsection{Crossed by-passable binary delay register}

\begin{figure}[H]
\center
\includegraphics[scale=.45]{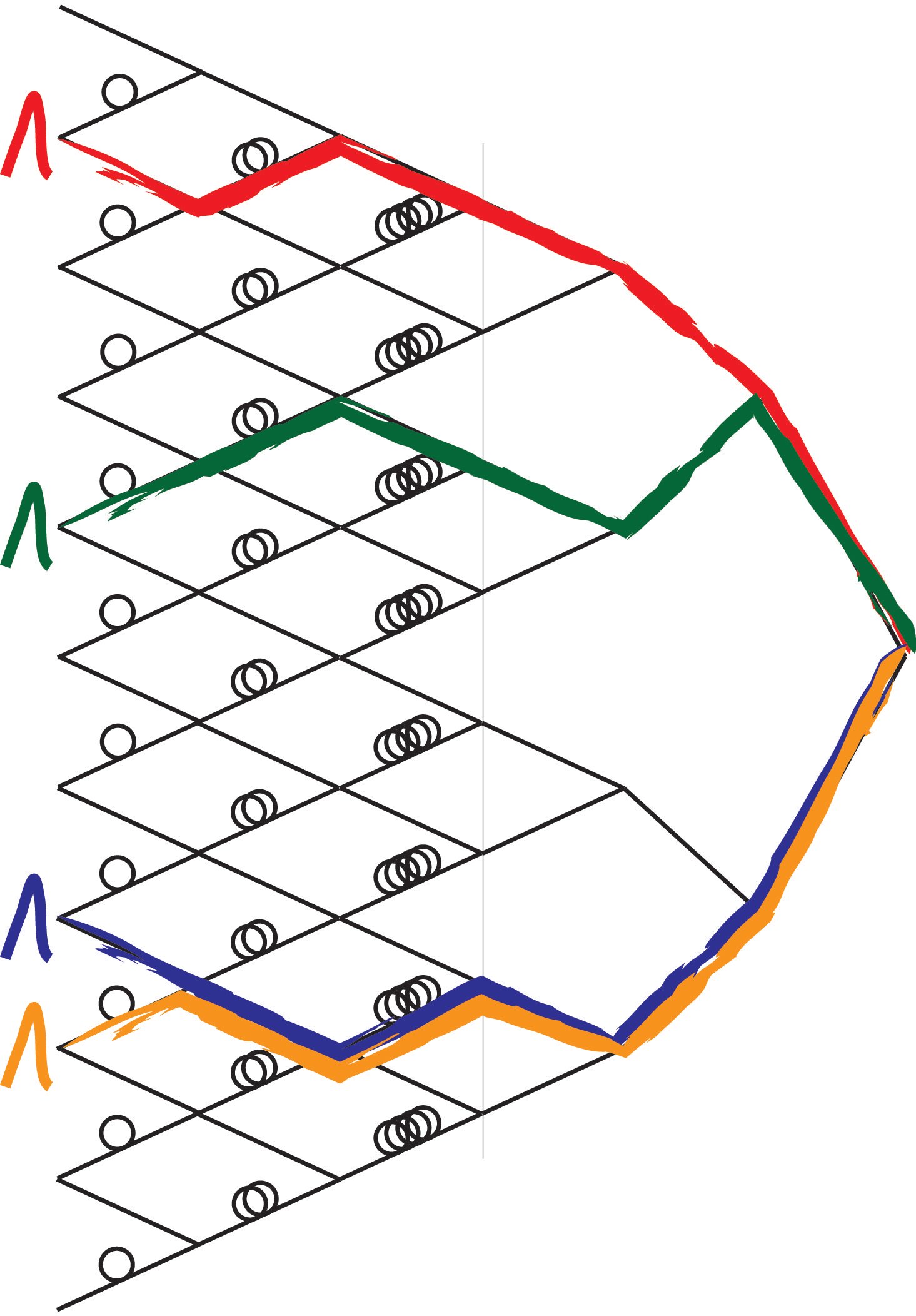}
\caption{$0$-$7T$ delay register paths. Delays: red: $2T$, green: $3T$, blue: $4T$, yellow: $5T$.}
\label{design_exemple11_use}
\end{figure}

The crossed by-passable delay register (CBDR) structure is the heart of the invention. It distributes the photons in a train of period $T$. The CBDR can achieve photon delays between $0T$ to $(2^N-1)T$ with $N$ number of delay steps. Thus, a 3-step register can delay photons from $0$ to $7T$ (Fig.~\ref{design_exemple11_use}).\\

In a given architecture, due to the asymmetry of the structure, control of boundary photons is more or less limited as shown on Table~\ref{table_1}. It is not an important problem if we choose good actuation rules (see \emph{fast-to-slow top-to-down driving} in the next sections) or if we simply remove one or two boundary downconverters.

\begin{table}[ht]
\centering
\begin{tabular}{c c c c}
\hline\hline
SPDC N\textordmasculine & Inaccessible states & SPDC N\textordmasculine & Inaccessible states \\ [0.4ex]
\hline
1 & 1T, 2T, 3T, 4T, 5T, 6T, 7T & N-2 & 0T\\ 
2 & 3T, 5T, 6T, 7T  & N-1& 0T, 1T, 2T, 4T\\
3 & 7T & N & 0T, 1T, 2T, 3T, 4T, 5T, 6T\\ [0.4ex]
\hline
\end{tabular}
\caption{Limitation of control of boundary downconverters for a $0$-$7T$ register structure.}
\label{table_1}
\end{table}


\subsection{Routing tree}
\label{routing_tree}

This structure simply drives the different delayed photons into a single output. The amount of switches increases in a logarithmic scale of the number of delayed single-photon inputs. It also allows to drive out of the structure the possible excess of photons.\\

\begin{figure}[H]
\center
\includegraphics[scale=.36]{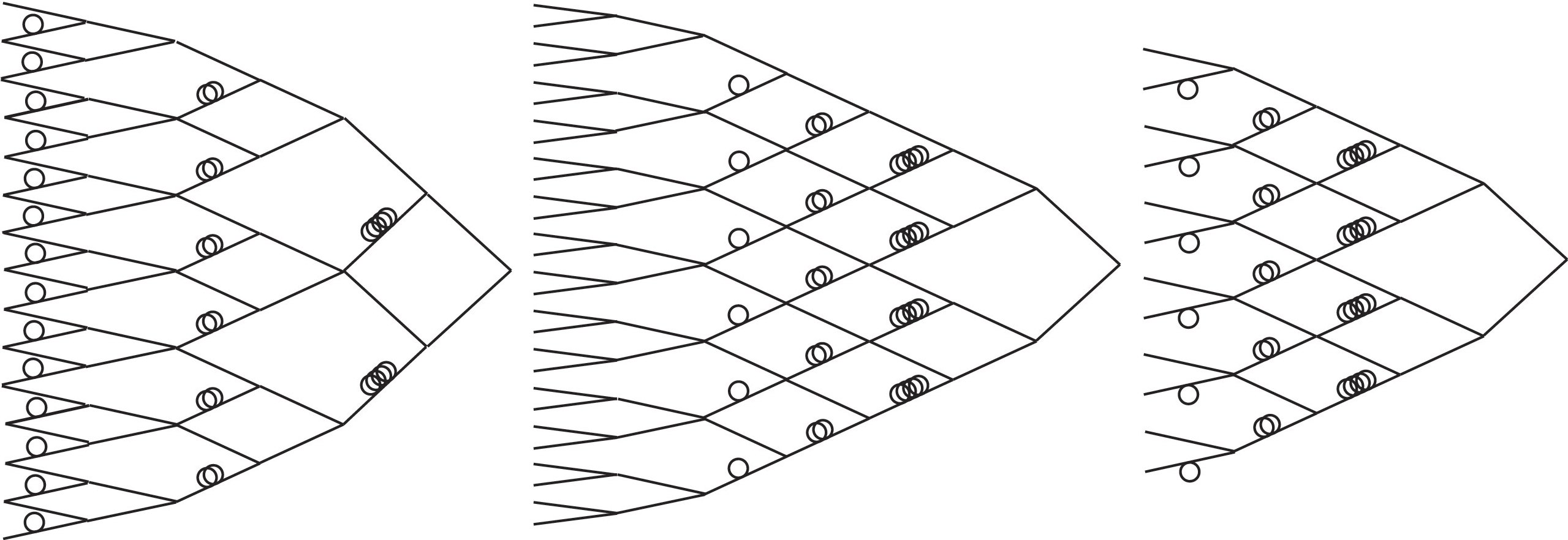}
\caption{Alternative structures. (left) Routing distributed in between delaying gives a more efficient integration on chip. (center) Routing before and after delaying is optimized for very low pumping power and high quality single-photon. (right) Reducing of one layer of switches by parity trade-off.}
\label{designs}
\end{figure}

For large structures and because of the low single-photon generation probability per downconverter (usually lower than 10\%), we can imagine different trees. It may be distributed before, after or in between the steps of the crossed binary delay register (Fig.~\ref{designs}). In this case, a more complex driving of photons needs to be established to avoid photon bunching (but more opportunities come to drive out concurrent or excess) or to consider it (using MZI switches in different configurations).

\section{Detailed functioning}

\subsection{Output rate up-conversion}

The CBDR structure and routing tree build a T-periodic photon train at each pumping pulse on the downconverter array. The rate of downconverter and the number of downconverters is matched in order to obtain enough photons to emit a train of several photons at the frequency of the pumping. The photon source frequency is thus a multiple of the pump frequency. With a $0$-$7T$ by-passable binary delay register, the longest photon train is 8. In this example, the maximal frequency multiplication is 8.\\

The maximal photon source frequency is fixed by the CBDR implementation to the first delay step T. The device frequency can be divided by skipping register steps and drive some photons out of the routing tree.\\

By trade-off on stability or multi-photon emission rate, the pumping period can be changed and lowered by several $T$, the frequency multiplication is thus lower.


\subsection{Stability enhancement}

Multiplying frequency needs a constant number of photons per pump pulse. Probabilistically, it needs in average much more photons than the frequency multiple. It is achievable by increasing again the downconverter number or pump power, but will lead to multi-photon emissions and large waste of photons.\\

\begin{figure}[H]
\center
\includegraphics[scale=.54]{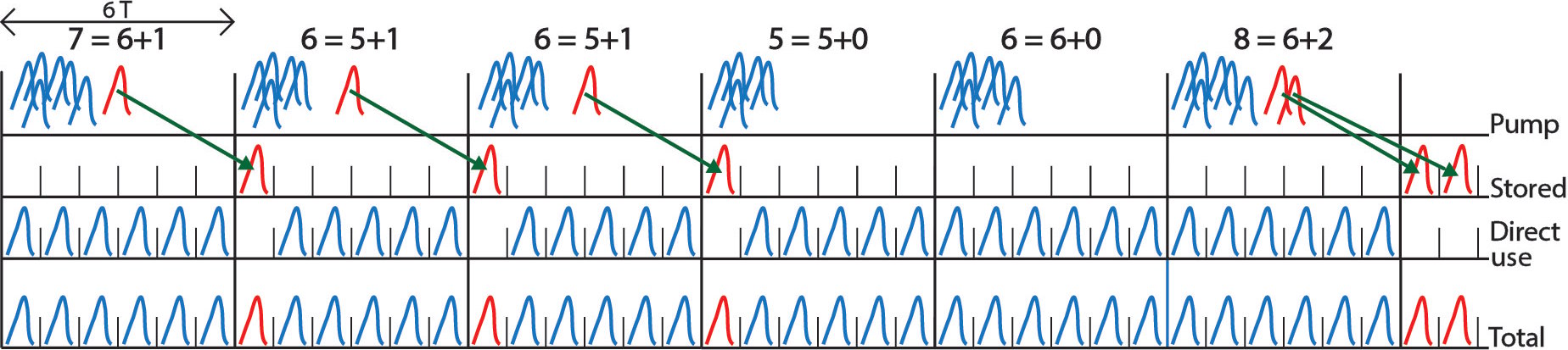}
\caption{Photon source frequency 6 times higher than pump rate. 1 or 2 photon excess (red) is stored and used at the beginning of the next cycle before the newly produced photons.}
\label{table_delay}
\end{figure}

Keeping the source frequency below the maximum frequency multiple make possible a \emph{temporary photon storage} of the excess photons into the longest delay lines. Photons can be delayed up to the next pump cycle and so will be emitted in the first place in this new pumping period. A little photon excess can be used during the next pulse, and a small lack can be corrected by the storage from the previous pulse (Fig.~\ref{table_delay}). This 1-cycle ahead memory allows to reduce the SPDC array average rate almost down to the frequency conversion multiple.

With the $0$-$7T$ register example and by lowering the frequency multiplication to 6, two photons can be stored for the next pulse cycle. 1 or 2-photon excess can be used during the next pulse, and 1 or 2-photon lack can be corrected by the storage from the previous pulse (Fig.~\ref{table_delay}). In these conditions, probability of lack of photon at the output drops significantly, since it only occurs when several successive low photon numbers are issued from the downconverter array.\\

The size of the downconverter array will be a factor of stability in addition to increase the total photon rate. Stability can once more be improved easily by adding a feedback to increase the pumping power for the next cycle to refill the temporary photon storage.


\subsection{Switching decisions}

The best strategy to operate the suggested device and avoid CBDR limitation problems or photon bunching problems is to adopt a \emph{fast-to-slow top-to-down} driving approach. It consists in allocating fastest photons (low delays) to slowest photons (high delays) from the top to the bottom of the CBDR structure.\\

In this operating mode, we reduce the need of inaccessible paths due to CBDR limitations and avoid any bunching of photons. Photons may share the same path, but at different times (Fig.~\ref{design_exemple11_use}).\\

With this approach, during a pumping cycle, each switch of the CBDR structure and the routing tree needs to be permuted \emph{at most once} and all in the same direction (figure out the permutations in the CBDR and in the routing tree switches needed to drive the photon train out on Fig.~\ref{design_exemple11_use}). The use of photon storage for stability enhancement slightly changes this property by delaying some permutations to the next pump cycle.


\section{Advantages to prior art}

\begin{itemize}
\item Frequency up-conversion of the laser pumpimg rate to the photon source output.
\item Tolerate low detector speed.
\item High stability of the output photon rate.
\item Possibility of feedback on the next cycle due to the efficient photon storage.
\item High controllability at fixed source frequency (pump power, pump frequency, power or frequency feedback on next pump pulse).
\item High control of the trade-off between frequency rise/multi-photon rate/output stability.
\item Few waste of photon.
\item Efficient use of multiplexing and customizable structure (SPDC number and CBDR level).
\item Easy up-to-down switching programming.
\end{itemize}


\chapter{Detailed analysis of the SPDC based single-photon source with output rate up-conversion multiplexed architecture}
 
\section{Case analysis of emission probability}

The following table shows various configurations of SPDC based single-photon source adapted with different error rates.

\begin{figure}[H]
\center
\includegraphics[scale=0.5]{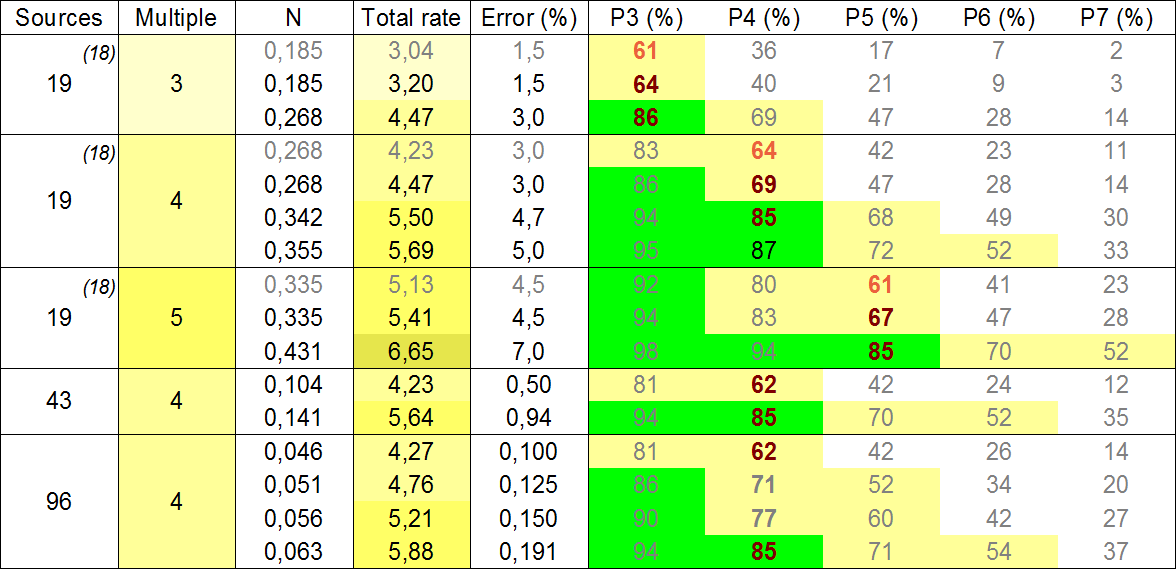}
\caption{Optimization of minimum photon emission probabilities and total error rate in function of number of sources, frequency multiple, mean photon number per module. First row of 19 sources tables with 18 sources considered for border limitation. Method: case analysis.}
\label{prob_graph}
\end{figure}

\section{Simulation}

A simulation tool has been realized with Maple. It assumes negligible photon loss inside the structure, considering a perfect waveguide and switching network. The following graphs show the absolute rate of lack error and multi-photon error of different configurations and parametrs of SPDC based single-photon source.\\

Each point has been simulated on 100,000 working cycles (on which multiplication factor for frequency must be applied to obtain the total photon number in ideal case) using the standard random tool of Maple. The simulator is also able to manage different power feed back responses (\emph{boost} and \emph{turbo-boost} modes).

\subsection{Working parameter variation}

The following Fig.~\ref{100a4power}~and~\ref{100a2_7} display the working characteristics of a large structure of 100 SPDC modules used without boosting with variation of frequency multiplication factor and pump power.

\begin{figure}[H]
\center
\includegraphics[scale=1]{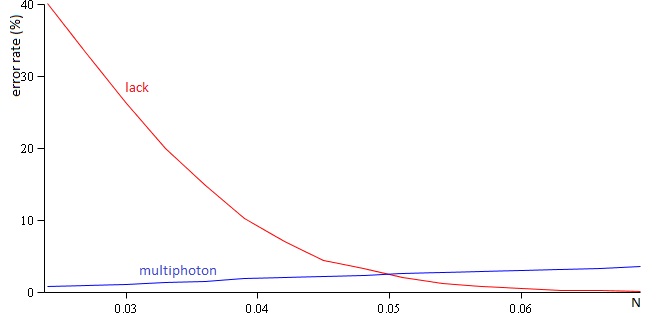}
\caption{Error rates of a 100 SPDC structure with frequency multiple of 4 in function of the pump power (or N).}
\label{100a4power}
\end{figure}

The results of the pump power variation for large structure as 100 SPDC modules show very low error rates and good possibilities of error trade-off. It gives an excellent device with high output gain.\\

The optimized point, with equivalence of lack and multi-photon error rates, is obtained at N $\approx$ 0.05. The source quality is then around 95~\% with 2.4~\% multi-photons and 2.4~\% lacks. No equivalent theoretical prior art SPDC source could achieve this quality at this rate.

\begin{figure}[H]
\center
\includegraphics[scale=1]{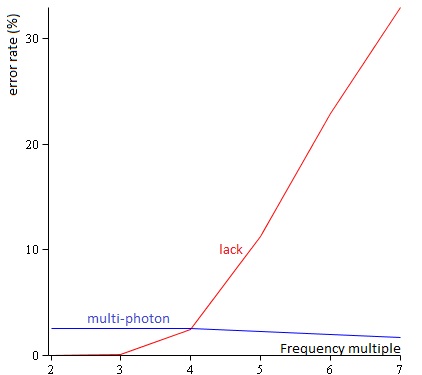}
\caption{Error rates of a 100 SPDC structure in function of the frequency multiple (optimized power at N = 0.049 for a gain output rate of 4).}
\label{100a2_7}
\end{figure}

We note that it is very easy to reduce the lack by changing the output multiple rate of the device. In the case of a variable pump frequency as feed-back, lack could be drastically reduced while keeping an equivalent multi-photon error rate.\\

The multi-photon rate depends linearly of the pump power. We note that on the figures, particularly on Fig.~\ref{100a2_7}, the pump power is fixed but the multi-photon error rate is not constant in the region of large lack. It is due to the absolute and homogeneous units used in this report, contrary to the relative \emph{multi-photon within emitted photon} unit used in most of papers.\\

\begin{figure}[H]
\center
\includegraphics[scale=0.8]{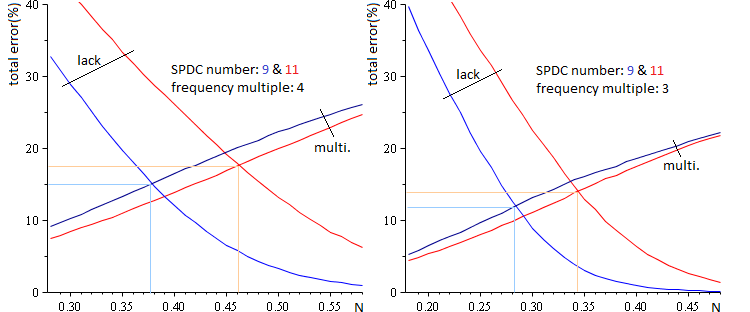}
\caption{Error rates of 9 and 11 SPDC structures at multiple of 3 (right) and 4 (left) in function of the pump power.}
\label{9_11a3_4}
\end{figure}

The Fig.~\ref{9_11a3_4} shows several working configurations of very small structure (9 and 11 SPDC modules) with optimized points for output gain of 3 and 4. For these small structures, I preferred to keep working at high output gain that lowering the error rates as it is a strong advantage of this device on the alternative solutions.\\

\subsection{Structure size}

\begin{figure}[H]
\center
\includegraphics[scale=1]{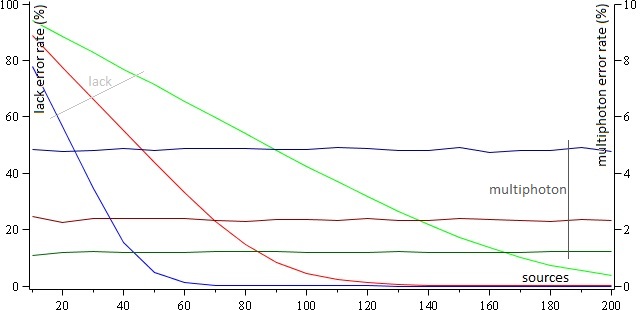}
\caption{Lack error rate and relative multi-photon error rate of SPDC structure depending on size for three pump powers (green: N=0.025, red: N=0.05, blue: N=0.10), with frequency multiple of 4.}
\label{10_200a4power1242}
\end{figure}

In Fig.~\ref{10_200a4power1242}, we analyze the variation error rates in function of the number of SPDC module of the structure.

%
%

\addcontentsline{toc}{chapter}{Bibliography}

\end{document}